\documentclass[aps,pra,twocolumn]{revtex4-1}
\usepackage{graphicx} 
\usepackage{tabularx}
\def\gapx{\lower 2pt \hbox{$\buildrel>\over{\scriptstyle{\sim}}$\ }}
\def\lapx{\lower 2pt \hbox{$\buildrel<\over{\scriptstyle{\sim}}$\ }}
\def\ph2{{\it p}-H$_2$}
\def\beq{\begin{equation}}
\def\eeq{\end{equation}}
\def\Am3{\AA$^{-3}$}
\begin{document}

\widetext
\title{Classical and quantum filaments in the ground state of trapped dipolar Bose gases}
\author{Fabio Cinti}
\email{cinti@sun.ac.za}
\affiliation{National Institute for Theoretical Physics (NITheP), Stellenbosch 7600, South Africa}
\affiliation{Institute of Theoretical Physics, Stellenbosch University, Stellenbosch 7600, South Africa}
\author{Massimo Boninsegni} 
\affiliation{Department of Physics, University of Alberta, Edmonton, Alberta, Canada T6G 2G7}
\date{\today}

\begin{abstract}
We study by quantum Monte Carlo simulations the ground state  of a harmonically confined dipolar Bose  gas  with aligned dipole moments, and with the  inclusion of a repulsive two-body potential of varying range.  
Two different limits can be clearly identified, namely a classical one in which  the attractive part of the dipolar interaction dominates and the system forms an ordered array of parallel filaments, and a quantum-mechanical one, wherein filaments are destabilized by zero-point motion, and eventually the ground state becomes a uniform cloud. The physical character of the system smoothly evolves from classical to quantum mechanical as the range of the repulsive two-body potential increases. An intermediate regime is observed, in which ordered filaments are still present, albeit forming different structures from the ones predicted classically;  quantum-mechanical exchanges of indistinguishable particles across different filaments allow phase coherence to be established, underlying a global superfluid response.
\end{abstract}

\maketitle
%
%
\thispagestyle{empty}

\section*{Introduction}
Spatially confined assemblies of  dipolar gases are the subject of ongoing experimental and theoretical investigation, mainly driven by the search for novel phases of matter that may arise as a result of the long-ranged, anisotropic character of the interparticle interaction \cite{menotti}. In particular, the experimental achievement of Bose-Einstein Condensation of  atomic systems with large magnetic moments \cite{griesmeier,lu,aikawa,depaz,ni,yan,takekoshi,park,balewski} suggests the intriguing possibility of identifying and studying experimentally phases simultaneously displaying ferromagnetism and superfluidity.
\\ \indent
The peculiar features of the interaction between two electric or magnetic dipoles bring about issues of stability of a three-dimensional (3D) Bose-Einstein condensate (BEC). Specifically, if all dipoles are aligned and the BEC is elongated in the same direction of the dipoles, then the dipolar interactions are mostly attractive, leading to the collapse of the BEC, much like in a generic Bose system with purely attractive interactions. This effect, predicted at the mean field level, was experimentally observed in a trapped assembly of chromium atoms \cite{chromium}. 
A different outcome was recently  reported \cite{drop}, however, in a similar experiment involving dysprosium (Dy), in which collapse was replaced by the formation of ordered arrays of droplets, reminiscent of the Rosenweig instability in  classical ferrofluids \cite{cowley,timonen}. In a subsequent experiment with erbium (Er), a smooth evolution of the system from a trapped BEC to a dense macrodroplet was observed, as the $s$-wave scattering length of the atomic interaction was  varied \cite{chomaz}.
\\ \indent
It was initially proposed that the failure of the Dy BEC to collapse and the ensuing appearance of crystals of droplets, each comprising a significant fraction of all particles in the gas, may be underlain by interactions involving triplets of atoms\cite{xi,bisset}; experimental evidence \cite{barbut} and theoretical arguments \cite{petrov,santos1,santos2,bisset2}, however, point to both effects arising  from the mere presence of a repulsive core at short distance in the interaction among pairs of atoms.
Indeed, recent quantum Monte Carlo (QMC) studies have yielded evidence of both the stability of a trapped dipolar BEC against collapse \cite{saito}, as well as the appearance of droplets \cite{macia}, if a short  range repulsion is added to the pair-wise dipolar interaction. 
\\ \indent
The physical origin of such a repulsive term could be different, depending on the physical system that one is considering; certainly any naturally occurring atomic or molecular interaction features a hard core repulsion at short distance arising from Pauli exclusion principle, which prevents electronic clouds of different atoms from overlapping spatially. The effective hard core diameter in that case is of the order of few \AA, i.e., much smaller than the typical value of the characteristic dipolar length (see below) in the majority of current experiments with cold dipolar atoms or molecules. Significantly greater ranges could be achieved, e.g., by means of the Feshbach resonance \cite{feshbach}.
\\ \indent
An interesting question that remains open is whether this system may be a potential candidate for the experimental realization of a supersolid phase, namely one possessing superfluid properties while simultaneously, {\em spontaneously} breaking translational invariance \cite{rmp}. A number of theoretical suggestions have been made of specific cold atom systems and settings, wherein this elusive phase of matter may be unambiguously observed, for example with Rydberg atoms \cite{cinti10,saccani1,saccani2,jltp,maucher,msc,nc}; experimentally, evidence of novel phases displaying density ordering and superfluidity has been recently reported for atomic BECs featuring spin-orbit interactions \cite{macche1}, or coupled to the modes of optical cavities \cite{macche2}. 
\\ \indent
We report  here results of an extensive theoretical investigation, based on QMC simulations, of the ground state of a harmonically confined  assembly of dipolar spin-zero bosons. We adopted the same microscopic model utilized in Ref. \cite{macia}, including a hard core repulsive term and with parameters of the harmonic confining potential chosen to be close to those of typical experimental set-ups.  We explore the different physics that emerges as the number of particles in the harmonic trap and the range of the short-distance repulsion are varied. 
\\ \indent
The results of the simulations show that, while one can not properly speak of ``phase'' of a finite system, there is a clear evolution from a classical to  a quantum regime, as the effective diameter of the hard core repulsion is increased. Specifically, in the limit of vanishing hard core diameter the energetics of the system is dominated by the potential energy, namely the attractive part of the dipolar interaction; the  breakdown at low temperature  of the uniform gas into parallel filaments forming ordered structures can thus  be easily understood in classical terms. In this regime, quantum-mechanical exchanges of indistinguishable particles across filaments are absent.
\\ \indent
On the other hand, as the range of the hard core repulsion approaches a significant fraction (i.e., a few tenths) of the characteristic length of the dipolar interaction, quantum mechanical effects of zero-point motion, as well as exchanges of indistinguishable particles, first cause filaments to thicken, becoming more similar to prolate droplets, then to merge, forming ordered structures that are different from those predicted classically, and eventually to disappear altogether, the ground state  becoming a uniform BEC (i.e., a single, dilute cloud), consistently with the experimental findings of Ref. \onlinecite{chomaz}.
In the intermediate regime, henceforth referred to as ``of quantum filaments'', exchanges of particles across different filaments underlie a global superfluid response, similarly to what observed in supersolid droplet crystals \cite{cinti10}.
\\ \indent
It need be mentioned that our results are in qualitative and quantitative disagreement with those reported in Ref. \cite {macia}; most significantly, whereas it is contended therein that the ground state of the system features ordered arrangements of filaments (in this regime, prolate droplets may be more appropriate a term) in a {\em narrow} range of values of the relevant interaction parameter (which, as stated above, is simply the characteristic diameter of the short distance repulsive core), we show here that, for a sufficient number of particles, the breakdown into filaments  occurs whenever this parameter is small compared to the natural length of the dipolar interaction. Indeed, as mentioned above such a breakdown {\em inevitably} occurs in the classical limit (i.e., of vanishing hard core diameter).
Although the results shown here pertain to a specific  choice of geometry of the confining trap, one similar to that adopted in most experiments, the basic physics illustrated here ought not depend on the details of the confinement.
\\ \indent
The remainder of this article is organized as follows: in Section \ref{model} we introduce the microscopic model, while in Sec. \ref{method} we offer details of the calculation carried out in this work; in Sec. \ref{res} we illustrate our results, outlining our conclusion in Sec. \ref{conclusions}.
\section{Model}\label{model}
We consider an ensemble of $N$ Bose particles of spin zero,  mass $m$ and dipole moment $D$,  moving in three dimensions in the presence of a harmonic confining potential centered at the origin. All dipole moments are aligned, pointing in the positive $z$ direction.
Henceforth, we express all lengths in terms of the characteristic length of the dipolar interaction, namely $a\equiv mD^2/\hbar^2$,
whereas $\epsilon\equiv (D^2/a^3)=\hbar^2/(ma^2)$ is the unit of energy and temperature (i.e., we set the Boltzmann constant $k_B=1$).\\ \indent
The  Hamiltonian of the system in dimensionless
units reads as follows:
\begin{equation}\label{u}
\hat H =-\frac{1}{2}\sum_i\nabla^2_i+\frac{1}{2\xi^4}(x_i^2+y_i^2+\gamma^2 z_i^2)+\sum_{i<j}U({\bf r}_i,{\bf r}_j)
\end{equation}
where ${\bf r}_i\equiv (x_i,y_i,z_i)$ is the position of the $i$th particle, $\xi$ is the confining length of the harmonic trap, and $\gamma$ is a trap elongation factor in the $z$ direction; in this work, we choose the same values of these parameters utilized in Ref. \cite{macia}, namely $\xi=1.2$ and $\gamma=4.0$, i.e., the trap is squeezed in the $z$ direction. This choice is aimed at rendering the geometry quantitatively close to that of most experiments. 
\\ \indent
The interaction potential $U$ between any two particles in the trap is given by
\begin{equation}
U({\bf r},{\bf r}^\prime)=U_{sr}(|{\bf r}-{\bf r}^\prime|)+V_d({\bf r},{\bf r}^\prime)
\end{equation}
where $V_d$ is the dipolar interaction which, since all dipole moments are aligned in the $z$ direction, is readily expressed as follows:
\begin{equation}
V_d({\bf r},{\bf r}^\prime)=\frac{1}{|{\bf r}-{\bf r}^\prime|^3}\ \biggl (1-\frac{3(z-z^\prime)^2}{|{\bf r}-{\bf r}^\prime|^2}\biggr ),
\end{equation}
while $U_{sr}$ is the short-distance repulsive potential, which in this study is modeled as in Ref. \cite{macia}, namely
\begin{equation}\label{sr}
U_{sr}(r)=\biggl (\frac{\sigma}{r}\biggr)^{12}
\end{equation}
The expression (\ref{sr}) is  that of a potential steeply rising for distances $\lesssim \sigma$; it cannot be given a rigorous theoretical justification, and is adopted here  for computational convenience. Its purpose is merely that of imparting a short-distance repulsion to the pair-wise interaction, thereby preventing particles from ``falling" onto one another (with the ensuing collapse of the system) as a result of the dipolar attraction that they experience as they approach one another along the $z$ direction. The parameter $\sigma$ does not, to our knowledge, have a direct relationship with the experimentally measured scattering length, but can nevertheless be roughly regarded as a hard core diameter.
\begin{figure}[ht]
\centering
\includegraphics[width=\linewidth]{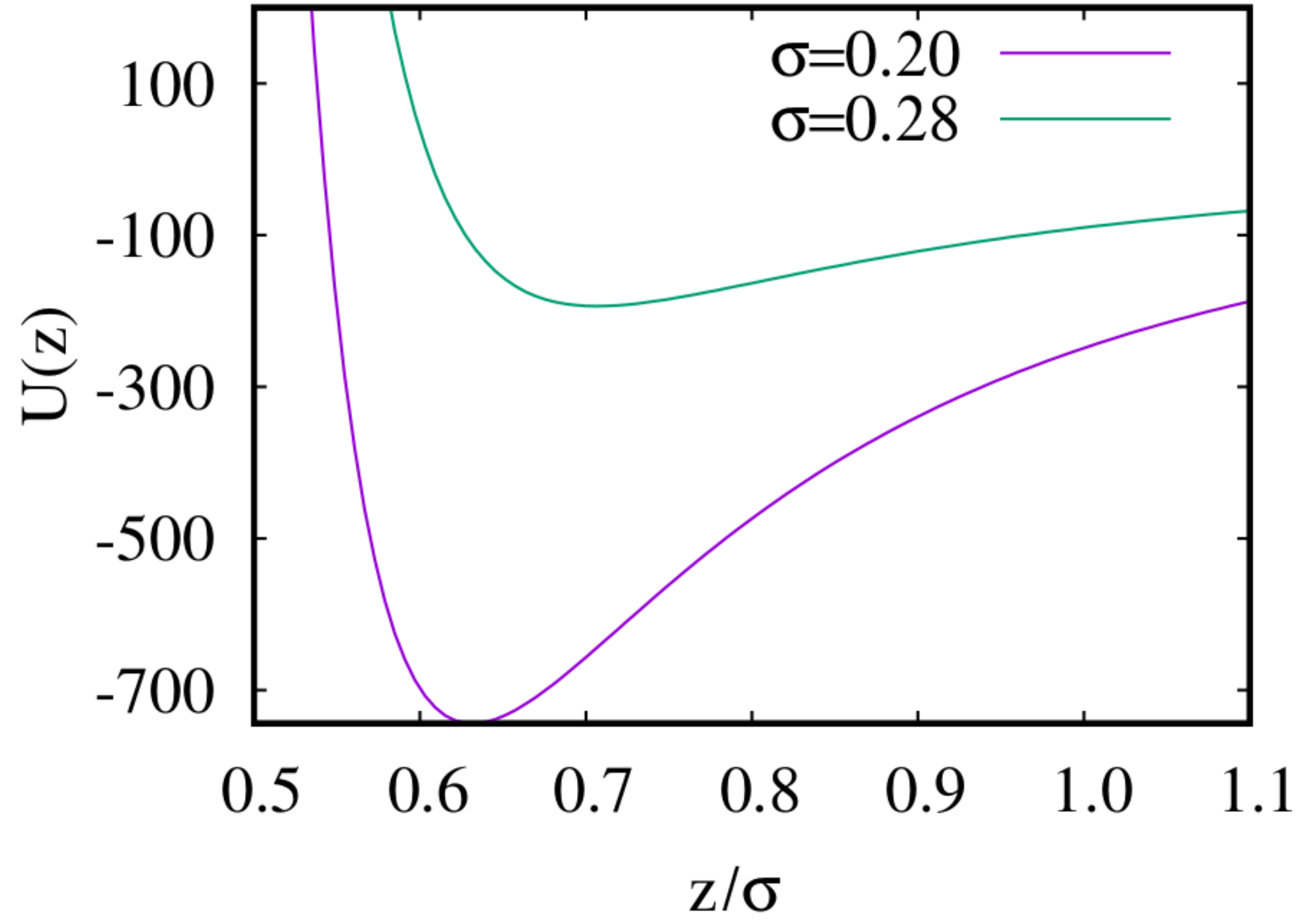}
\caption{{\rm Color online.} Interaction potential (in the energy unit $\epsilon$ defined in the text) between two particles approaching one another head-on, along the direction of alignment of the dipole moments ($z$), for two values of the hard core parameter $\sigma$. Potential energy values are in units of $\epsilon$.}
\label{f1}
\end{figure}

\begin{figure}[ht]
\centering
\includegraphics[width=\linewidth]{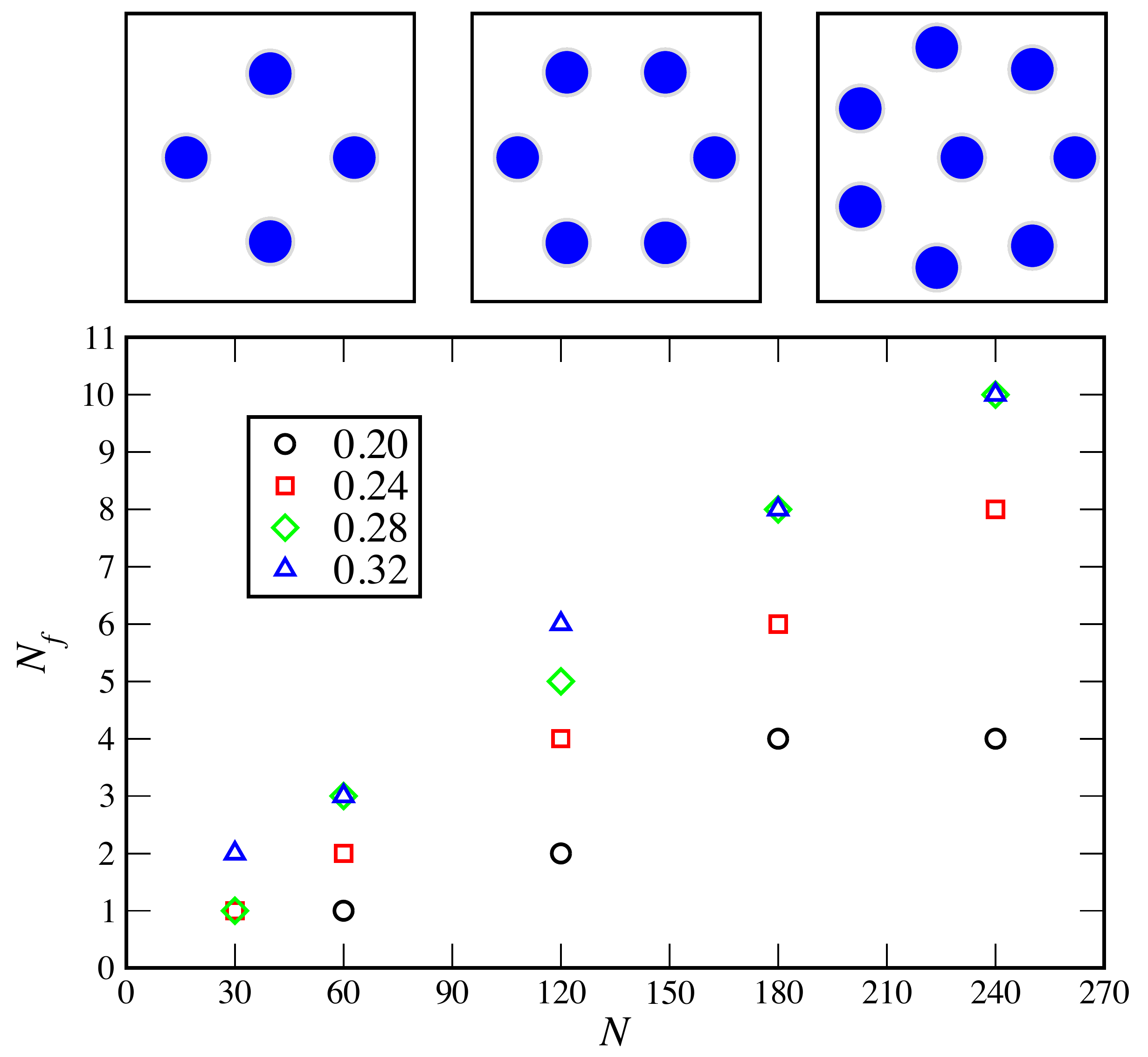}
\caption{{\em Color online.} {\em Top:} Typical crystalline structures formed by filaments (top view). {\em Bottom:} Number of filaments in the classical ground state of model (\ref {u}) as a function of total particle number $N$, for different values of $\sigma$.}
\label{f2}
\end{figure}

The dipolar potential $V_d$ is anisotropic; it is purely repulsive (decaying as $1/r^3$) for two particles laying on a plane perpendicular to the direction of dipole alignment, and it is most attractive for two particles approaching one another head-on along that same direction. Figure \ref{f1} shows the resulting potential on including the repulsive part (Eq. \ref{sr}). As $\sigma\to 0$, the attractive well becomes increasingly deep, leading to the conclusion that the potential energy will dominate in this limit, and the behavior of the system will be classical.
\\ \indent
The potential energy minimum is achieved by ``piling up'' particles along the $z$ direction, i.e., forming a one-dimensional (1D) filament with an inter-particle separation approximately given by $\sigma^{4/3}$; in the presence of a confining potential, however, as $N$ grows confinement along $z$ makes it energetically advantageous for the system at some point to start forming separate filaments. Two finite, rigid parallel filaments can be  straightforwardly shown to exert a repulsive force on one another, and therefore the classical ground state of the system for large $N$ consists of filaments arranged in ordered structures, similar to those formed by, e.g., dipolar particles in a two-dimensional harmonic trap \cite{me13}. A few examples of such structures are shown in the top part of Fig. \ref{f2}.

For a given trap geometry, as well as number $N$ of particles and value of $\sigma$, it is a   straightforward exercise to identify the classical ground states, by direct minimization of the potential energy. Figure \ref{f2} shows the number $N_f$ of filaments in the classical ground state of (\ref{u}) as a function of $N$, for the trap geometry adopted here. As we can see, $N_f$ is an increasing function of $N$, more rapidly so the greater $\sigma$, as fewer and fewer particles can be  accommodated in a single filament due to the increased 1D lattice constant.

As mentioned above, the classical picture should hold in the limit $\sigma\to 0$, as the attractive well of the anisotropic, pair-wise potential grows deeper. On the other hand, as $\sigma$ increases and becomes of order one (i.e., comparable to the dipolar length $a$), the well is shallow and one can expect quantum zero-point motion, as well as exchanges of indistinguishable particles \cite{pollet}, to destabilize the classical filaments in favor of a uniform cloud or perhaps different ordered structures. In order to investigate this scenario, and in particular to account in full for quantum-mechanical effects, we adopted first principle numerical (QMC) simulations.
\section{Methodology}\label{method}
The low temperature phase diagram of the system described by Eq.  (\ref{u}) as a function of the total number $N$ of particles in the trap, as well as of the hard core parameter $\sigma$, has been studied in this work by means of  first principles numerical simulations, based on the continuous-space Worm Algorithm \cite{worm,worm2}.  Since this technique is by now fairly well-established, and extensively described in the literature, we shall not review it here. We used a variant of the algorithm in which the number of particles $N$ is fixed \cite{mezz1,mezz2}.
Details of the simulation are mostly  standard, one difference with respect to previous works is that we used here  the so-called ``primitive'' approximation for the short imaginary time ($\tau$)  propagator, which requires a greater number of time slices than more elaborate propagators (e.g., the fourth order one \cite{jltp2}) but which  we found more convenient, due to its simplicity, when dealing with an anisotropic interaction. Obviously, all of the results presented here are extrapolated to the $\tau\to 0$ limit; in general, we found numerical estimates for structural and superfluid properties of interest here, obtained with a value of the time step $\tau\sim 10^{-3}\epsilon^{-1}$ to be indistinguishable from the extrapolated ones, within the statistical uncertainties of the calculation.
\\ \indent 
Because we are mainly interested in the physics of the system in the $T\to 0$ limit,  we generally report here results corresponding to temperatures $T$ sufficiently low to regard them as essentially ground state results, mainly because further lowering the temperature does not result in significant qualitative or quantitative changes. This is typically achieved at  higher temperature in the classical (i.e., $\sigma\to 0$) limit, but in general we observed that simulations carried out at a temperature $T\sim\epsilon$ can be regarded as representative of the ground state of the system.
\\ \indent
We carried out simulations of systems typically comprising few hundred particles, 1024 being the largest size considered here; for the trap geometry utilized, this corresponds to a density range between few tens and few hundreds $a^{-3}$. We typically started our simulations from many-particle configurations corresponding to the classical ground states for the chosen values of $N$ and $\sigma$. This is the same protocol normally adopted when simulating, e.g., quantum crystals, which consists of placing particles initially at lattice sites consistent with the crystalline structure of interest. We have also used as starting configuration a single filament at the center of the trap, as well as a high temperature, gas-like configuration. 
As we are primarily interested in the filament structure of the system, physical conclusions are mainly drawn based on a visual inspection of many-particle configurations generated in the course of sufficiently long simulation runs. 
\\ \indent
Regarding the superfluid response, in principle it can be evaluated using the well-known ``area'' estimator \cite{sindzingre}, appropriate for a finite system such as the one considered here. In practice, the filament structure of the system renders the value of the superfluid fraction comparatively low in the supersolid phase (not unexpectedly \cite{jltp}), and therefore we rely on a more indirect criterion, namely we monitor the frequency of cycles of permutations of identical particles involving a number thereof significantly greater than that in a single filament. 
While there is no quantitative connection between permutation cycles and the superfluid fraction \cite{mezzacapo08}, and although it is possible for individual, quasi 1D filaments to possess an individual superfluid response (e.g., in the Luttinger sense), a global superfluid phase requires exchanges of particles across filaments (see, for instance, Ref. \onlinecite{cinti10}).

\section{Results}\label{res}
We begin with a qualitative illustration of the physical trend occurring as the value of the parameter $\sigma$  is increased.
\begin{figure}[h]
\centering
\begin{tabular}{cc}
\includegraphics[width=0.46\columnwidth]{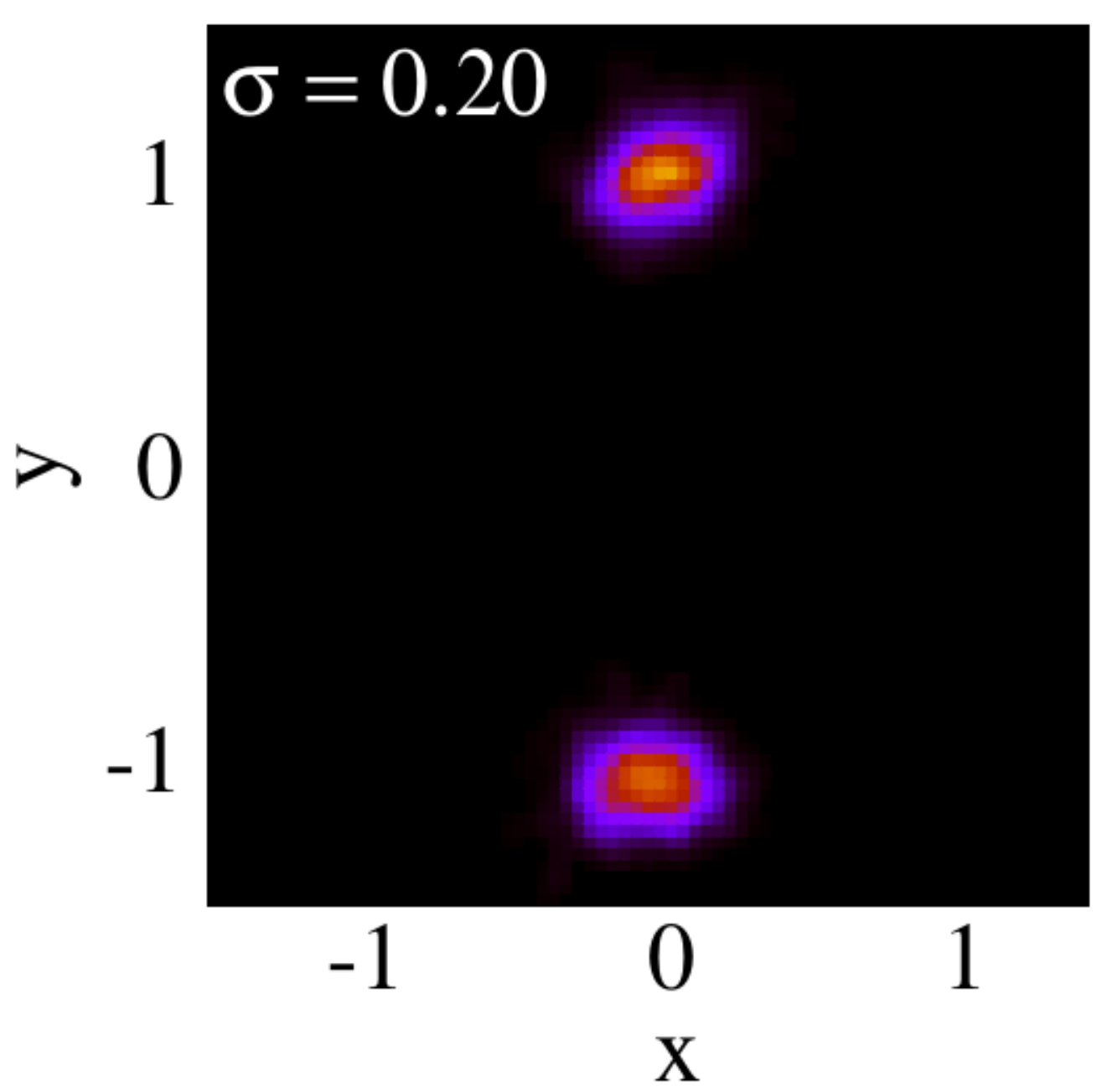}&
\includegraphics[width=0.46\columnwidth]{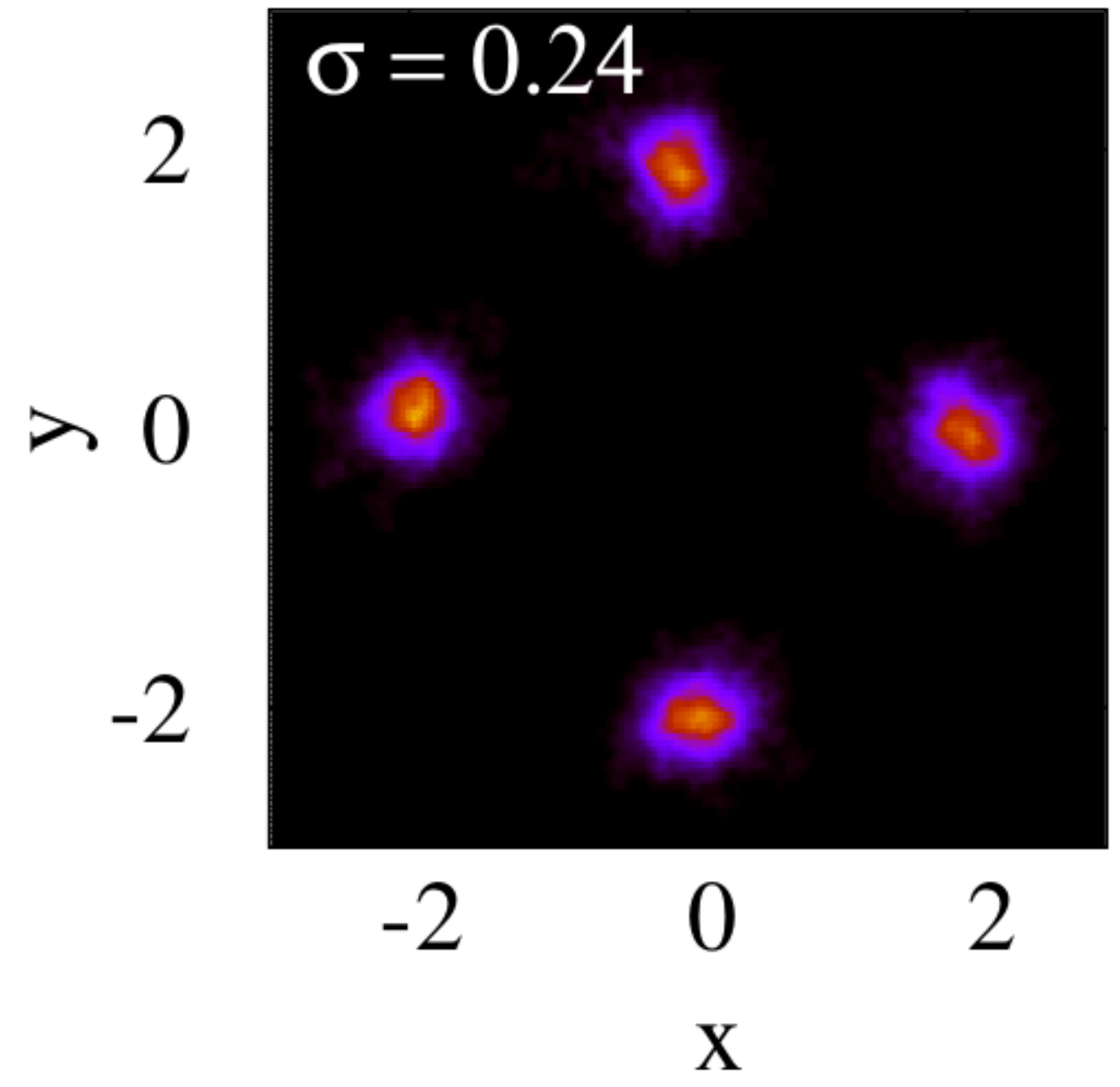}\\
\includegraphics[width=0.46\columnwidth]{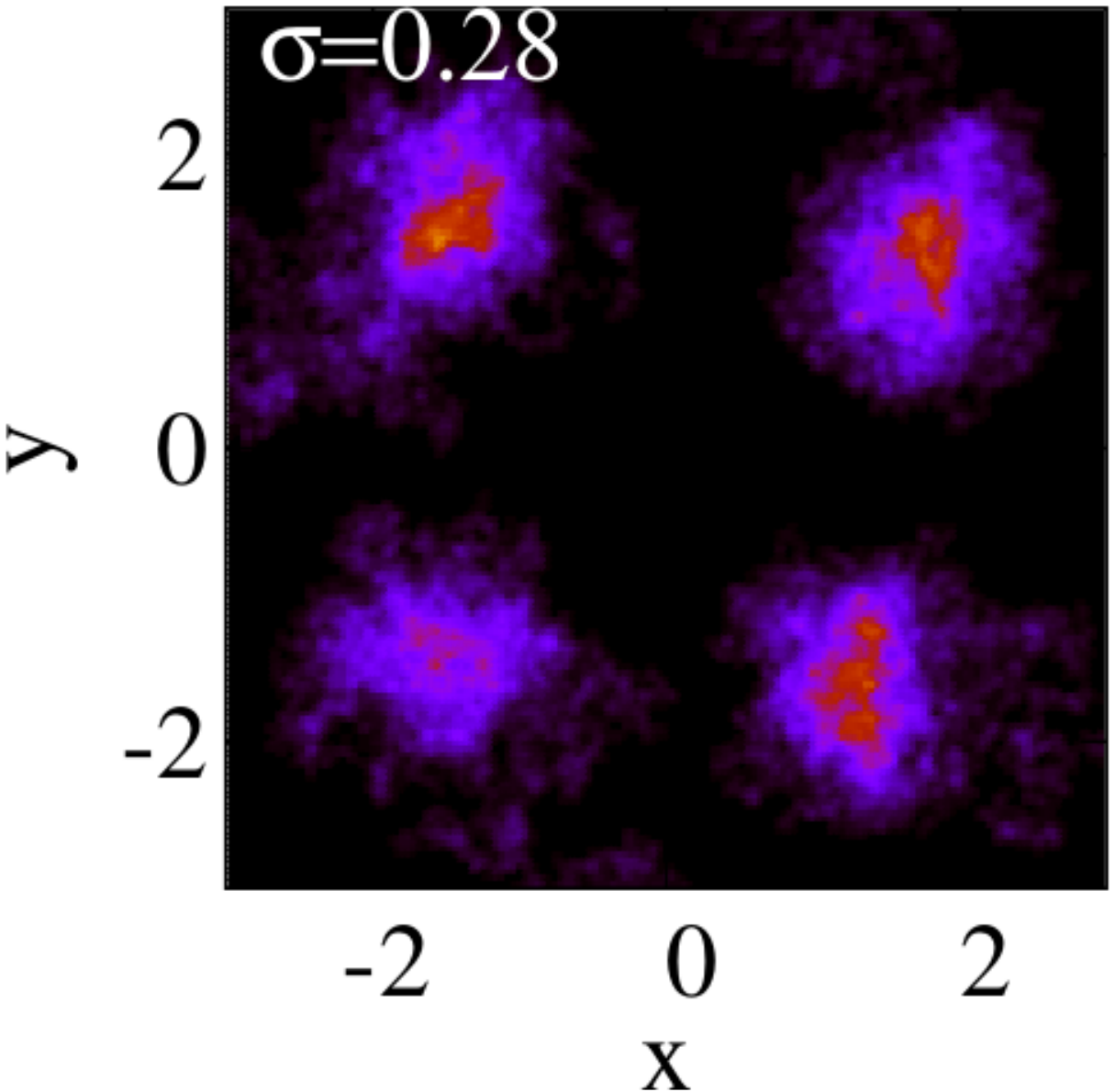}&
\includegraphics[width=0.46\columnwidth]{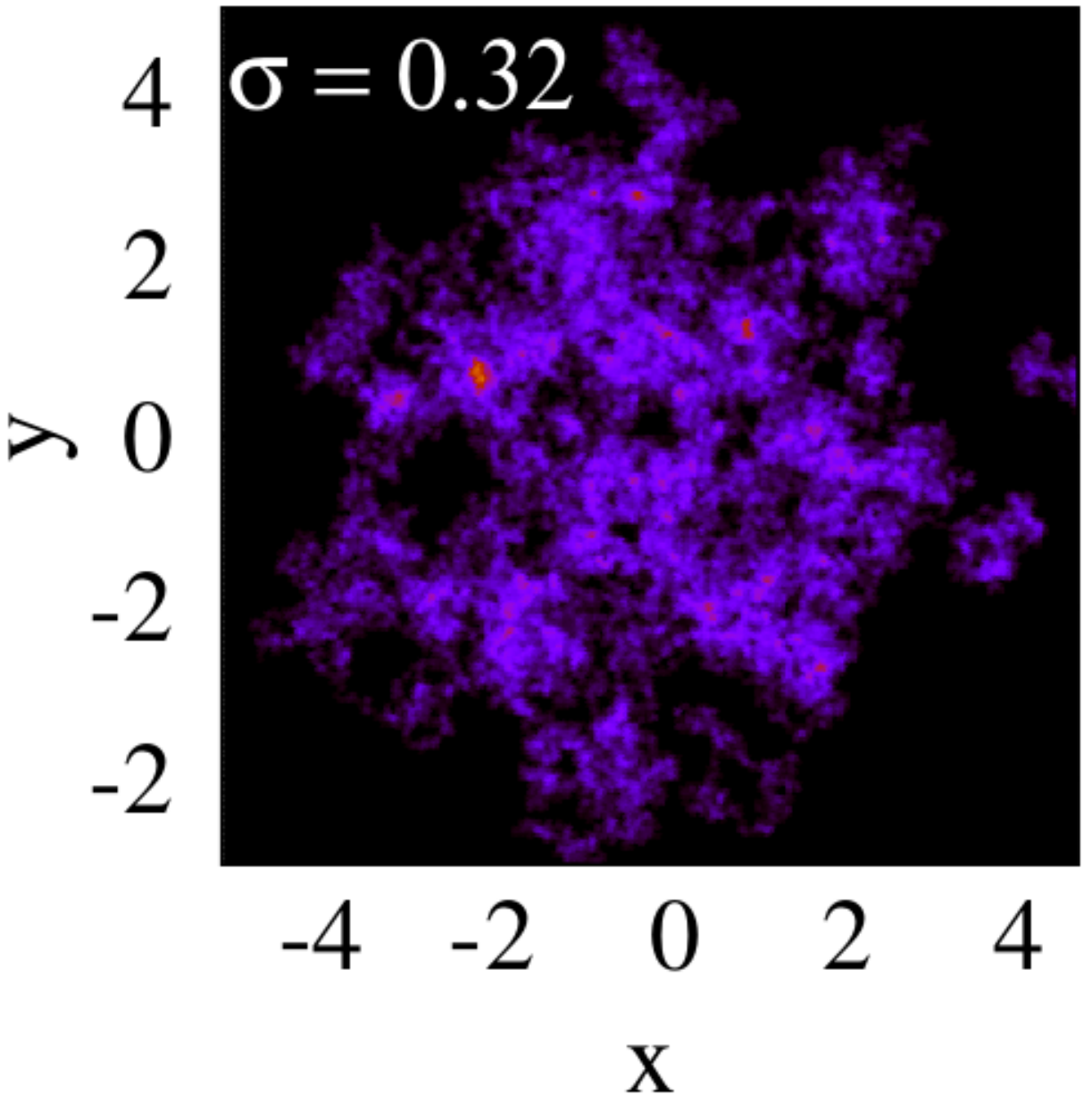}\\
\end{tabular}
\caption{{\it Color online}. Density maps for a system of $N$=120 particles (top view), for four different values of the parameter $\sigma$. All lengths are all in units of $a$ (see text). 
}\label{f3}
\end{figure}
\begin{figure}[h]
\centering
\begin{tabular}{cc}
\includegraphics[width=0.46\columnwidth]{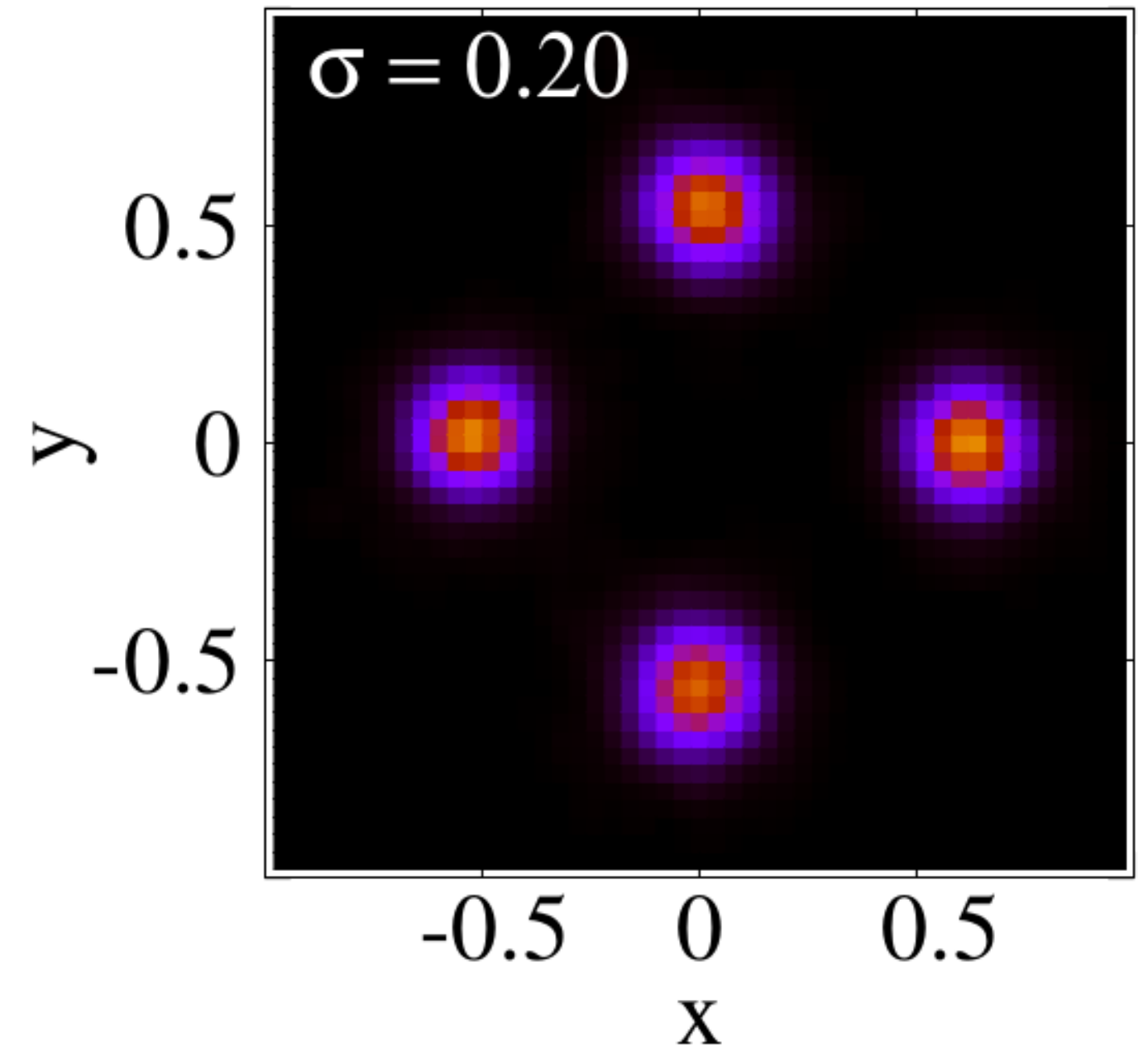}&
\includegraphics[width=0.46\columnwidth]{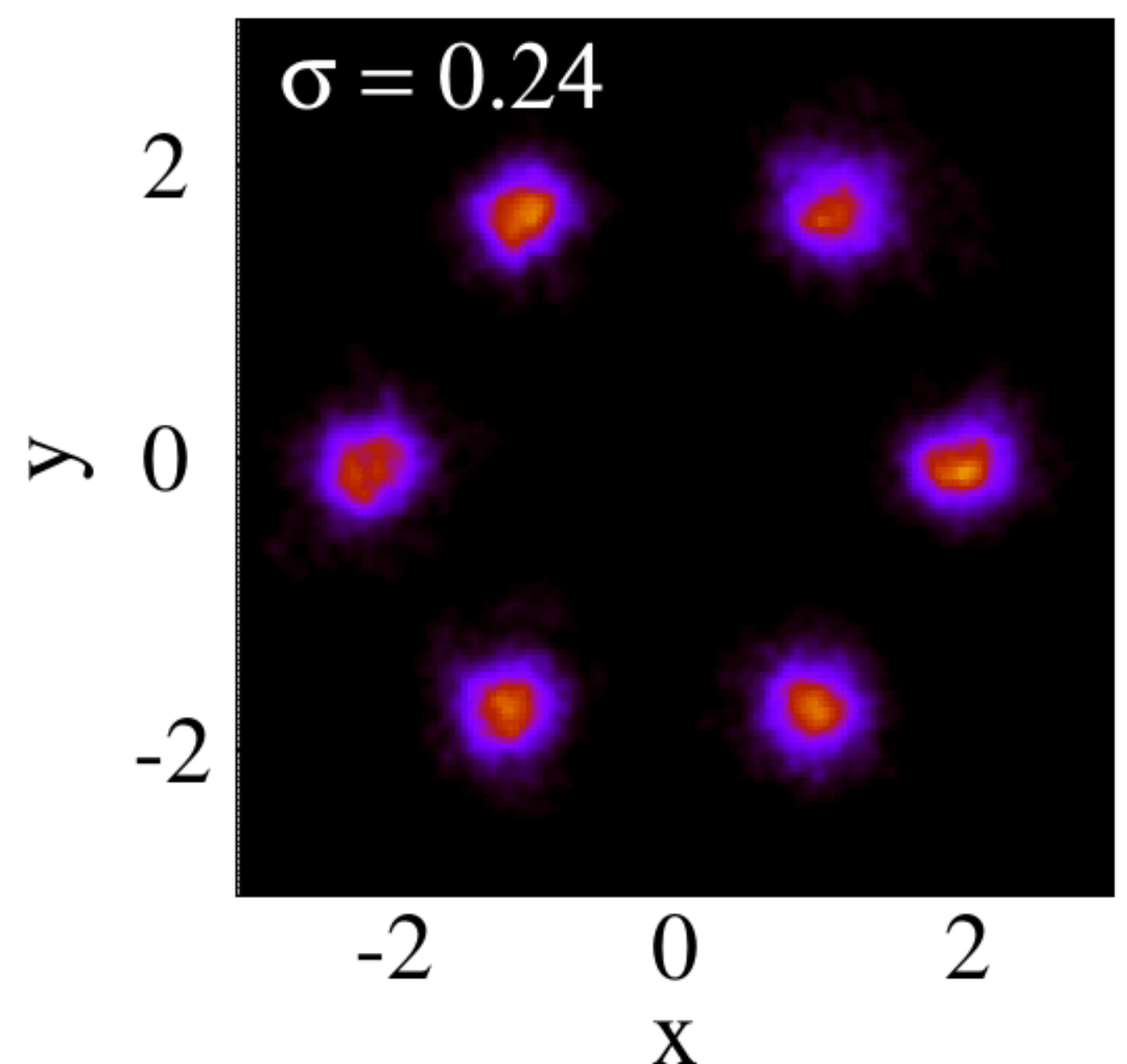}\\
\includegraphics[width=0.46\columnwidth]{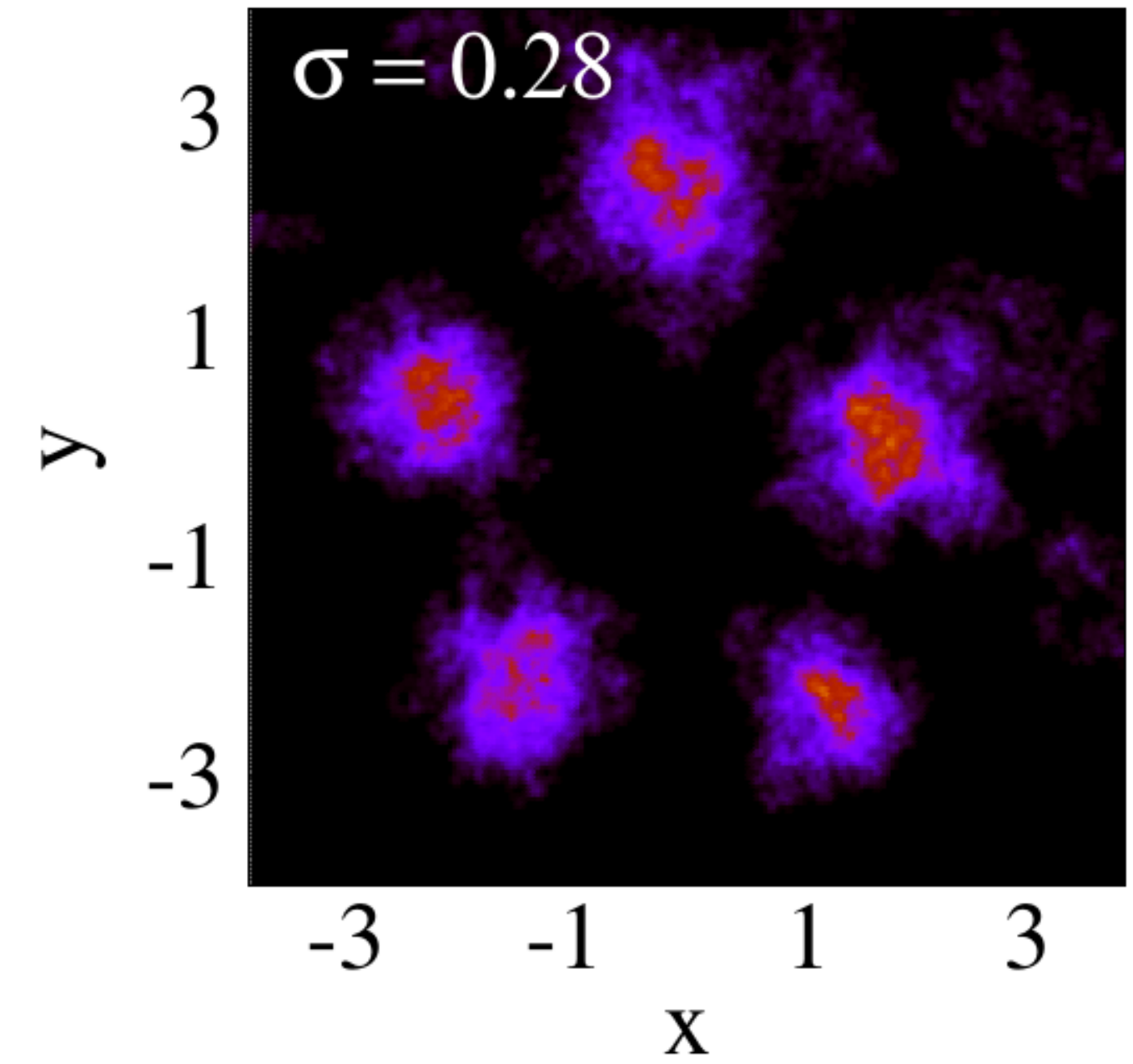}&
\includegraphics[width=0.46\columnwidth]{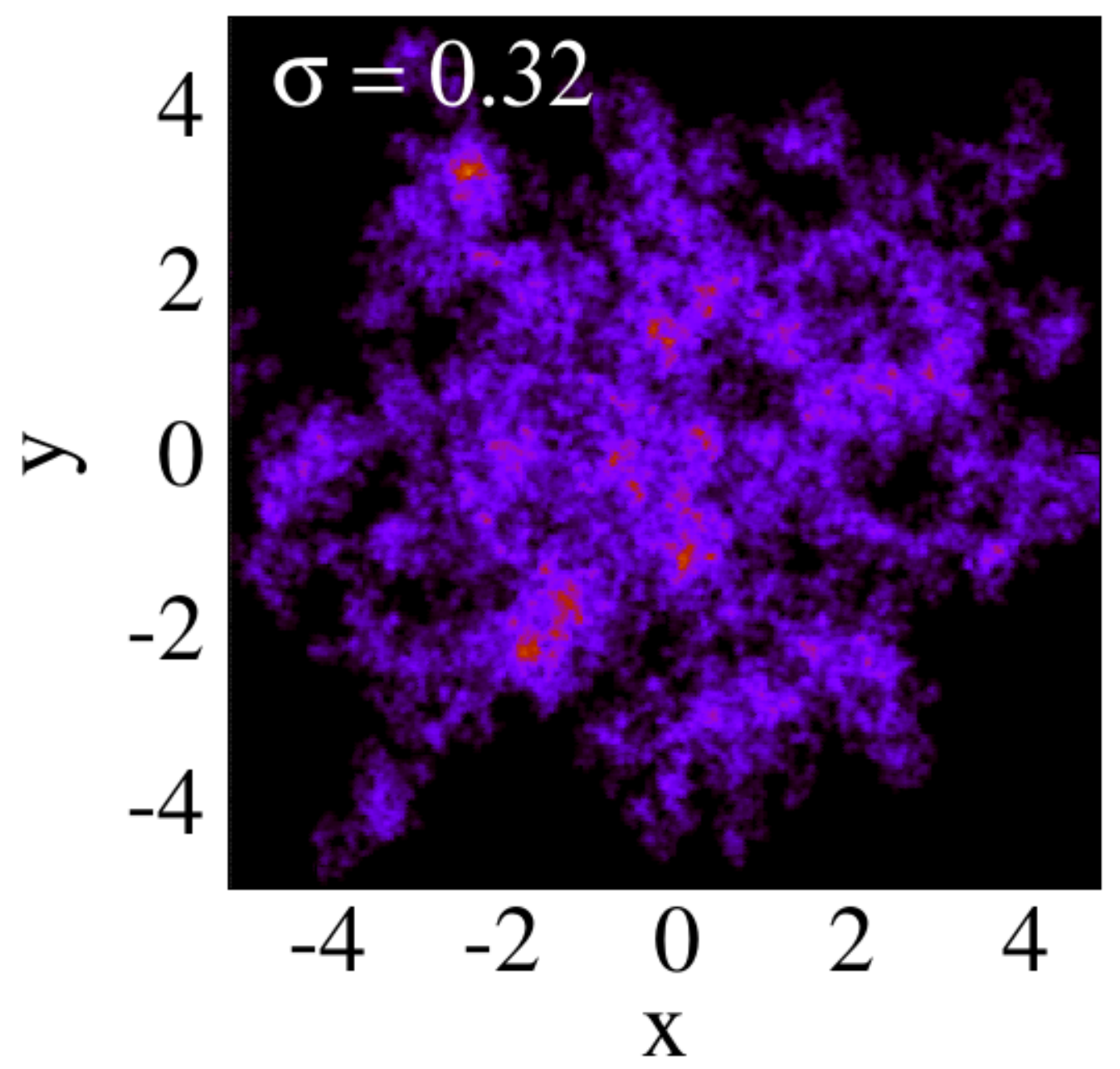}\\
\end{tabular}
\caption{{\it Color online}. Same as Fig. \ref{f3} but for $N$=180 particles. 
}\label{f4}
\end{figure}
Figures \ref{f3} and \ref{f4} show two-dimensional density maps (integrated over the $z$ direction) for a system comprising $N=120$ and $N=180$ particles, for different values of the hard core parameter $\sigma$, namely 0.20, 0.24, 0.28  and 0.32. These maps are obtained from instantaneous many-particle configurations generated during a sufficiently long simulation. They are physically equivalent to all other configurations generated in the course of the same run for the specific conditions considered (e.g., others differ at the most by an overall rotation), and are therefore presented as statistically representative. The temperature of the simulations is in all cases low enough that the system may be regarded in the ground state; specifically, it is $T=5 \epsilon$ for $\sigma=0.20$, whereas $T=\epsilon$ for all other cases. We have, however, carried out simulations down to $T=0.5 \epsilon$ and ascertained that the physics remains the same in all cases.
\\ \noindent
A filament structure is clearly identified in three out of the four cases, whereas for the largest value of $\sigma$ the system forms a uniform cloud. As we shall see, however, there are substantial differences between the physical behavior at $\sigma \lesssim 0.24$, and that observed in the range $0.24 \lesssim \sigma\lesssim 0.3$.
\\ \indent
For values of $\sigma\lesssim 0.24$, we find that the simulated system remains very close to its starting configuration, which, as mentioned above, corresponds to the classical ground state. Quantum-mechanical zero point motion causes the  filaments to expand slightly in the transverse direction, acquiring a relatively small diameter (e.g., approximately 0.02 for $\sigma=0.1$), as illustrated in Fig. \ref{f3} for the two lower values of $\sigma$. Exchanges among particles in the same filament become relatively important at low temperature as $\sigma$    approaches a value close to $\sim 0.2$, but do not significantly affect the structure of the filament itself. On the other hand, particle exchanges across filaments are not observed in this regime, to which we refer as ``classical''.  
\\ \indent
In such a regime, filaments {\em always form}; we also observe that if the simulation is started from a different, ordered filament arrangement, e.g., including a greater number of filaments than the classical ground state, such a configuration remains stable, albeit its energy is greater than that obtained if the classical ground state is taken as a starting point. This can be interpreted as evidence of long-lived ordered filament configurations above the ground state.
\\ \indent
As $\sigma\to 0$, the density of particles in a single filament grows, i.e., more and more particles are needed to form more than one filament. For example, for a value of $\sigma=0.1$, the lowest considered in this work, the ground state of the system consists of a single narrow filament along the axis of the trap, for $N$ at least up to 1024. This observation is consistent with the results of the QMC simulation carried out by Saito \cite{saito}, based on a different model for the short-range repulsion, but with a choice  of the equivalent hard core parameter well into the ``classical'' regime (a quantitative comparison of the results obtained here with those of Saito would require that $\sigma$ be set to a value less than 0.1).  On the other hand, it was reported in Ref. \cite{macia}, based on a ground state QMC calculation making use of the same microscopic Hamiltonian adopted here, that filaments (prolate droplets)  only form in the narrow $0.24<\sigma <0.28$ range, which is clearly in disagreement with the arguments furnished and the results obtained here. We come back to this point below.
\\ \indent
As the hard core parameter $\sigma$ grows above a value $\sim 0.24$, the physics of the system changes qualitatively. Quantum zero-point motion and permutations of indistinguishable particles causes filaments to melt and/or merge, and for $\sigma\gtrsim 0.3$ the ground  state is a uniform cloud (100\% superfluid in the $T\to 0$ limit, the superfluid response uniform across the cloud as we have ascertained through the computation of the superfluid density using the area estimator \cite{mezzacapo08}), with no remnant of filaments, as shown in panel ({\rm d}) of Figures  \ref{f3} and \ref{f4}. There exists, however, an interesting intermediate regime around $\sigma=0.28$, wherein the system displays an ordered filament structure that is different from the classical one.
\\ \indent
Consider for definiteness the case $N=120$, $\sigma=0.28$, shown in panel ({\rm c}) of Fig. \ref{f3}. Four lumps can be clearly identified; although the simulation was initially started from the classical ground state, including {\em five} filaments (as shown in Fig. \ref{f2}), we observe these filaments to merge in a single cloud, which then divides into the four prolate droplets shown in Fig. \ref{f3}. Remarkably, this is consistently observed in this system, at sufficiently low $T$ ($\lesssim \epsilon$), regardless of the initial configuration chosen. Indeed, four droplets form even if the simulation is started from a single filament at the center of the trap, or, e.g., three filaments, or from a disordered, high temperature initial configuration; for example, the result shown in Fig. \ref{f3} for $\sigma=0.28$ pertains to a simulation in which particles were initially placed on a single filament. We therefore conclude that the presence of four droplets is an {\em intrinsic} quantum-mechanical signature of the ground state of the system, for this number of particles and this value of $\sigma$. These quantum droplets are not pinned at classical locations, but rather are observed to meander in the course of the simulation; their relative positions, however, remain fairly constant. The same behavior is observed in the simulation of the system comprising $N=180$ particles, for this value of $\sigma$, which forms five droplets at low $T$ irrespective of the starting configuration.
It should be noted that in this regime, to which we refer as ``quantum", exchanges of identical particles are not restricted to individual droplets, but extend across different droplets. 
\\ \indent 
The number of droplets in the ground state is consistently {\em less} than that predicted classically, as a result of consolidation arising from quantum exchanges, which results in an effective attraction among particles obeying Bose statistics. For example, in the case $\sigma=0.28$ and $N=180$ the number of droplets in the ground state is five, as opposed to the eight predicted classically (see Fig. \ref{f2}).
Altogether, the formation of droplets in this regime is strikingly reminiscent of the exchange induced crystallization already observed in a system of soft core bosons \cite{pohl}.

\begin{figure}[ht]
\centering
\includegraphics[width=\linewidth]{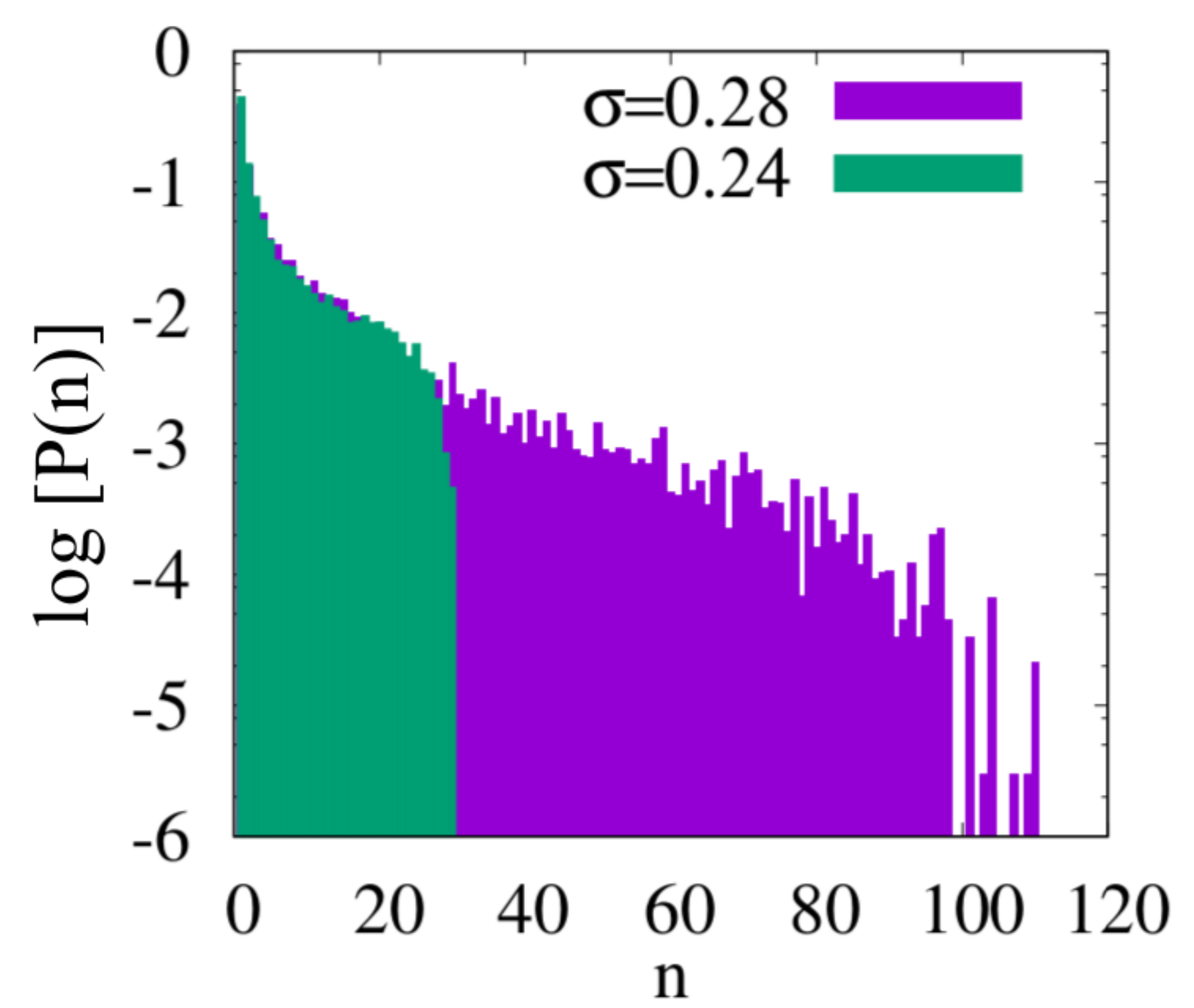}
\caption{{\em Color online.} Logarithm (base 10) of frequency of occurrence of cycles comprising $n$ particles, at low $T$ (see  text), for  a system of $N$=120 particles and for two different values of the parameter $\sigma$.}
\label{f5}
\end{figure}

As mentioned in the introduction, one of the interesting questions about this particular system is whether its low temperature phase diagram may include a {\em supersolid} phase; specifically, one may imagine a superfluid phase consisting of ordered filaments, each one individually superfluid, wherein phase coherence is established through particle exchanges across filaments. The results obtained in this work show that such a phase does not occur in what we refer to as the classical regime (i.e., $\sigma\lesssim 0.24$), wherein filaments are spaced apart and particle exchanges only take place within filaments. Even though individual filaments can be quasi 1D superfluids, the phase as a whole cannot acquire global phase coherence. On the other hand, in the quantum filament regime (roughly $0.24 < \sigma < 0.3$), particle exchanges across filaments are observed, leading us to make the conclusion that a supersolid phase indeed exists in this regime.
\\ \indent
To illustrate this point, Figure \ref{f5} shows the (logarithm of the) computed frequency $P(n)$ of occurrence of permutation cycles involving $n$ particles ($1\le n \le N$), for a system of $N$=120 particles. The distributions shown pertain to two different values of $\sigma$, namely 0.24 (in this case, the temperature is $T=\epsilon$)  and 0.28 ($T=0.5\epsilon$). The important point here is that the ground state features the same number of filaments (four) in both case, and the qualitative features of the two distributions do not change at lower temperatures. Now, while the $P(n)$ for $\sigma=0.24$ abruptly falls to zero for $n > 30$, which is the number of particles in a single filament, that for $\sigma=0.28$ features cycles involving almost all particles in the system.
As mentioned above, the presence of cycles involving all particles in a single filament for $\sigma=0.24$ makes it likely that filaments are superfluid (again, in the Luttinger sense, as these are quasi 1D objects); however, it is for $\sigma\sim 0.28$ that a supersolid phase is predicted, as the system still features an ordered array of filaments while at the same time displaying exchanges across filaments. Consistently, the superfluid response, evaluated with the area estimator, yields a finite value of approximately 20\% at $T$=0.5 \cite{note}.
This is the same physics observed in the supersolid cluster crystal phase \cite{cinti10,jltp}. On the other hand, for $\sigma \le 0.24$ the character of the system at low $T$ is that of an array of independent quasi 1D superfluids.

\section{Discussion}\label{conclusions}
In this work we carried out an extensive theoretical investigation of the low temperature (ground state) properties of a harmonically confined assembly of dipolar bosons, with dipole moments all aligned. Our study is based on first principle numerical simulations, and we adopted the same theoretical description recently utilized in another study, which includes, besides the dipolar interaction, a short-range repulsive potential to prevent the system from collapsing. Although we have made use of a specific trap geometry, chosen to mimic as closely as possible that of most experiments, the results presented here are general enough that they should be independent of that aspect. As well, the specific way in which we have modeled the short-range repulsion ought not affect the most important qualitative and quantitative conclusions drawn here. Indeed, the main physical effect that we set out to explore, namely the formation of a ordered arrays of filaments,  appears observable in a rather wide range of physical conditions, considerably more extended than suggested in Ref. \cite{macia}.
\\ \indent
In particular, in the limit in which the range of the hard core short distance  repulsion is small compared to the characteristic length of  the dipolar interaction, the system forms filaments as a result of a purely classical effect, arising from the competition between the attractive part of the dipolar potential and the confinement. As this limit is the one relevant to most current experimental settings, the observation of droplets in experiments is not altogether surprising. The physical character  of the filaments evolve as the hard core  diameter increases, causing a weakening of the dipolar attraction. While for a sufficiently large value of the hard core diameter filaments melt, and the ground state of the system is a uniform cloud of particles, we have found that there exists an intermediate region that has no classical counterpart, wherein filaments form due to purely  quantum effects, namely exchanges of indistinguishable particles obeying Bose statistics. These filaments form different ordered structures from those predicted  classically, which can give rise to a supersolid phase, as demonstrated by the simulation results shown here, underlain by particle exchanges across filaments. Whether these physical conditions, chiefly a hard core repulsion of range approximately one third of the characteristic dipolar length, are experimentally achievable at this time, remains to be ascertained.
\\ \indent
Our results are at variance with those of Macia {\em et al.} (Ref. \cite{macia}), not only regarding the number of droplets that form in the ground state of the system for specific  values of $\sigma$ and $N$, but also, much more significantly, because they ostensibly failed to observe any filament phase in the ``classical'' region, where they most obviously should occur, based on the simple (potential energy) considerations outlined above. In their  study, they employed the $T=0$ Path Integral Ground State method \cite{sarsa,cuervo}, which is in principle capable of projecting the true ground state of a Bose system out of an initial, non-orthogonal trial wave function. It need be noted, however, that  the  trial wave function utilized in that study is just a constant, i.e., it has very small overlap with a ground state consisting of few, quasi 1D filaments; it is therefore conceivable that the projection algorithm may need an exceedingly long time to converge to the true ground state, especially in a regime dominated by the potential energy. This is akin to the convergence problem that one would face on attempting to extract a crystalline  ground state from a non-interacting variational {\em ansatz} (i.e., a constant trial wave function), something in principle possible but usually unfeasible in practice. More generally, although zero temperature methods prove useful in specific cases, the experience accumulated over the past two decades points to finite temperature techniques as being typically  more reliable, at least for Bose systems \cite{1dmb}. 

\section*{Acknowledgements}

This work was supported in part by the Natural Sciences and Engineering Research Council of Canada (NSERC). M. B. gratefully acknowledges the hospitality of the National Institute for Theoretical Physics (NITheP), Stellenbosch, where most of this research work was carried out. The authors wish to acknowledge useful conversations with A. Kuklov, T. Macr\'i and S. Moroni. Computing support of Westgrid is also acknowledged.


\begin{thebibliography}{99}
\bibitem{note}
The superfluid fraction ought not saturate to 100\% even at $T=0$, in a phase that simultaneously breaks rotational invariance.
\end{thebibliography}
\end{document}